\documentstyle[twoside,fleqn,espcrc2]{article}
\input epsf.sty

\title{The viscous disk of GRS~1915+105}

\author{T. Belloni\address{Astronomical Institute ``A. Pannekoek'',
	University of Amsterdam and Center for High-Energy Astrophysics,
	Kruislaan 403, NL-1098 SJ, Amsterdam, The Netherlands}
	\thanks{This work is partly supported by NWO under grant PGS 78--277},
        M. M\'endez$^{\rm a,}$\address{Facultad de Ciencias
	Astron\'omicas y Geof\'{\i}sicas, Universidad Nacional
	de La Plata, Paseo del Bosque S/N, 1900 La Plata,
	Argentina},
        A.R. King\address{Astronomy Group, University of Leicester,
	Leicester LE1 7RH, United Kingdom},
	M. van der Klis$^{\rm a}$,
        J. van Paradijs$^{\rm a,}$\address{Physics Department,
	University of Alabama in Huntsville, Huntsville, AL 35899, USA}
}
\begin{document}

\begin{abstract}
GRS~1915+105, one of the two known galactic microquasars, shows an
extremely complex variability in the X-ray band, comparable to no other
X-ray source in the sky. Making use of RXTE/PCA data, we have analyzed
the X-ray spectral distribution throughout the variability. We find
that all variations can be attributed to the rapid appearance and
disappearance of the inner region of an optically-thick accretion disk.
Since the time scale for each event is related to the maximum radius
of the disappearing region,
the difference in time structure is due to the time distribution of
such radii. The observed relation between the extent of the missing inner region of
the disk and the duration of an event is in remarkable agreement with the
expected radius dependence of the viscous time scale of the 
radiation-dominated region of an accretion disk.
\end{abstract}

\maketitle

\section{INTRODUCTION}

GRS~1915+105 is the first galactic object for which radio jets moving
at superluminal speed have been observed \cite{miro94}. Radio observations
have also led to the estimate of its distance ($\sim$12.5 kpc) and the
inclinations of the jets ($70^\circ$ from the line of sight).
The source is considered to host a black hole because of its high 
X-ray luminosity and its similarities to GRO~1655--40, whose dynamical
mass estimate indicate the presence of a black hole \cite{bail95}.
GRS~1915+105 is a transient source, in the sense that it has made its
appearance in the
X-ray sky in 1992 \cite{ctbl92}, but most likely never returned into 
quiescence since then. The Rossi X-ray Timing Explorer (RXTE) monitors
regularly GRS~1915+105 since the start of the mission. During this period,
the source has shown a remarkable richness in variability, ranging from 
quasi-periodic burst-like events, deep regular dips and wild oscillations,
alternated with quiescent periods \cite{gmr96,mgr97,cst97,tmb97a,tmb97b,tcs97}.
We present here the results of the analysis of RXTE data of GRS~1915+105 and 
propose an interpretation for the variability in the spectral parameters.
A more detailed discussion can be found in \cite{tmb97b}.

\section{X-RAY OBSERVATIONS}

RXTE observed GRS~1915+105 in many occasions.  Here we present the
results of the analysis of one of these observations, the one obtained
on 1997 June 18 starting at 14:36 UT and ending at 15:35 UT, as it
reproduces within one day most of the variability observed from this
source.  The upper panel of Figure 1 shows 1200~s of the
2-40~keV light curve.  It consists of a sequence of `bursts' of
different duration with quiescent intervals in between.  All bursts
start with a well-defined sharp peak and decay faster than they rise.
The longer bursts show oscillation (or sub-bursts) towards the end.

The complexity of the light curve seems to be untractable: the only
obvious comment that can be made is that it consists of high flux intervals
somehow alternated with low flux intervals. In order to quantify the
timing structure of the oscillations, we measured the length of the
`ups' and `downs', an unambiguous procedure given
the square-wave structure of the light curve. Excluding the fastest
oscillations where this measurement is not easy, we obtain durations
spanning from a few seconds to three minutes.

To study the evolution of the spectrum of GRS~1915+105 during these
oscillations between `up' and `down', we first
produced a color-color (C-C) diagram of the whole observation
A characteristic pattern emerged, pattern
which is also observable, with small differences, in all other observations
we analyzed, besides a few (see below). This strengthens the hypothesis that 
all the oscillations have the same nature.
An analysis of separate C-C diagrams for oscillations of different length
(see Figure 2) shows systematic differences. As it can also
be seen from the light curve, the `ups' are very similar to each other,
while the `downs' are different, in particular the longer the oscillation
the deeper (light curve) and harder (C-C) is the `down'. Both from the
light curve and the C-C diagram it appears that the short `low' periods
are similar to the last part of the long ones.
Since spectral
analysis at the required time resolution is not possible, in order
to transform the information in the C-C diagram into spectral parameters
we divided it into small regions.  For all the
populated regions in the diagram we accumulated 1 second time
resolution energy spectra in 48 energy bands.  We measured a background
spectrum from a blank sky observation, which we normalized to the
highest energy channels where the contribution of the source was
negligible.  We subtracted this background spectrum from each of the
source energy spectra.  We used the latest detector response matrix
available, and added a systematic error of 2\% to account for the
calibration uncertainties. For each of the regions in the C-C diagram
we fitted the data with a ``standard'' spectral model for BHCs,
consisting of a disk-blackbody (DBB) model and a power law, both
affected by
interstellar absorption.  To avoid problems due to the background
subtraction, and as we were only interested in the properties of the DBB
component, we limited our fits to energies below 30~keV.  Since both the
distance to this system and the inclination of the accretion disk are
known (we assume that the jet is perpendicular to the disk), we can
derive the inner radius of the accretion disk directly from the fits
without significant additional uncertainties.

This rather complicated procedure allowed us to obtain time histories for the
inner radius and the temperature of the disk 
(bottom two panels of Figure 1).  During `ups' the temperature
is above 2~keV and the radius is stable around 20~km.  During `downs',
the temperature drops to less than 1~keV and the radius increases.

We produced similar C-C diagrams for a number of other RXTE observations
of GRS~1915+105.  All the observations that we analyzed can be fitted in
the same manner, except that of 1996 June 16th \cite{gmr96} (and some
later observations in October 1997).  The
quasi-periodic bursts observed in many of the observations \cite{tcs97}
are consistent with repetitive short events like the ones described
here.

\begin{figure*}[htb]
\epsfysize=9cm
\vspace{9pt}
\hspace{2.8cm}\epsfbox{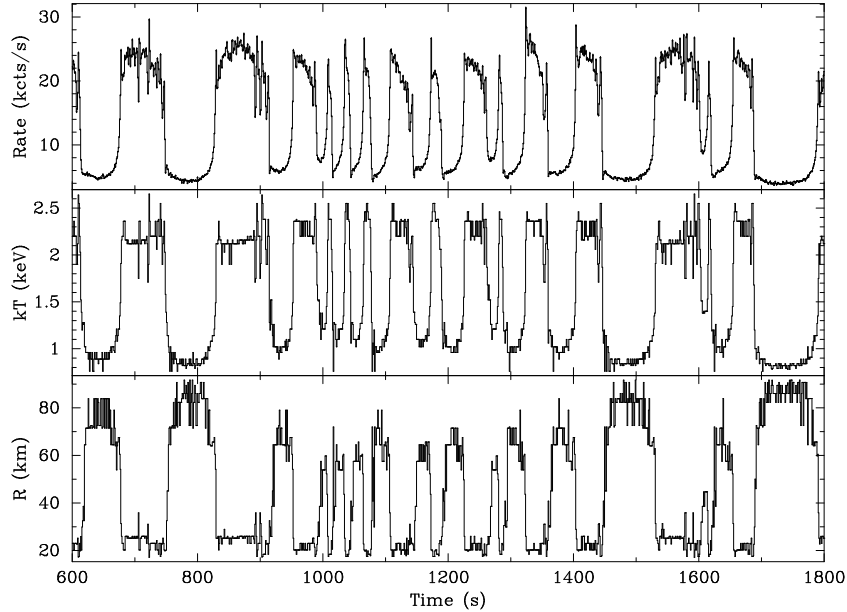}%
\caption{Upper panel: 2-40 keV PCA light curve. Time zero corresponds to
1997 June 18th, 14:36 UT. Middle and lower panels: corresponding inner radius
and temperatures (see text)}
\label{fig:1}
\end{figure*}

This analysis brought us to consider more seriously the `up' and `down'
paradigm: to be consistent with our previous work, we will call them
`outburst' and `quiescent' states. A sequence of quiescence and outburst
we call an `event'.

\section{DISCUSSION}

The main conclusion that can be derived from Figure 1 is that the 
inner radius of the accretion disk is not constant in time. If we
interpret the small and stable radius of $\sim$20 km observed during
outburst as the minimum stable orbit around a black hole 
(see \cite{tmb97a} for a more detailed discussion), this implies that 
during quiescence the central section of the disk disappears, i.e. a central
hole appears in the disk. The radius of this central hole varies between
events, ranging in this observation between 30 and 90 km (see Figure 1).
This can be interpreted within 
the model presented by us in \cite{tmb97a}, providing a unified picture
of the variability observed in GRS~1915. In \cite{tmb97a}, we modeled the 
large amplitude changes as emptying and replenishing of the inner 
accretion disk caused by a viscous-thermal instability. The small radius
observed during the quiescent period was identified with the innermost stable
orbit around the black hole, while the large radius during the burst phase
was the radius of the emptied section of the disk. The smaller
oscillations were interpreted as failed attempts to empty the inner disk.
As it can be seen from Figures 1 and 2, from this observation we find 
that all variations, from major events
like the ones described in \cite{tmb97a} to small oscillations observed at the
end of a large event, can be modeled in exactly the same fashion.
In this scenario, the ``flare state'' presented in \cite{tmb97a} is simply a
sequence of small events following a big one, similar to the small
oscillations in Figure 1.

Both the spectral evolution and the duration of the event are 
determined by one parameter only, namely the radius of the missing inner 
section of the accretion disk. It is natural to imagine that the bigger
the missing inner section of the disk, the longer it will take to 
re-fill it. Indeed, if we identify the re-filling time for each event with the
time spent in `quiescence', this is what is observed (Figure 3).
Following \cite{tmb97a}, we can naturally associate the
re-filling time of an event $t_q$ to the viscous time scale of the 
radiation-pressure dominated part of the accretion disk. This can
be expressed as $t_{\rm visc}=R^2/\nu$, where $\nu = \alpha c_S H$. 
using the expressions for the scale-height $H$ and sound speed
$c_S$ found in \cite{fkr92}, we obtain:
\begin{equation}
t_q\sim t_{\rm visc} = 
              30 \alpha_2^{-1} M_1^{-1/2} R_7^{7/2}\dot M_{18}^{-2}\ {\rm s}
\end{equation}
where $\alpha_2 = \alpha/0.01$, $R_7$ is the radius in units of $10^7$ cm, 
$M_1$ the central object mass in solar masses, and $\dot M_{18}$ is the 
accretion rate in units of $10^{18}$ g/s. Notice that even the largest
radii derived here are well within the radiation-pressure dominated part
of the disk (see Equation 2 in \cite{tmb97a}).
The line in Figure 3 represents the best fit to the data with a relation
of the form $t_{\rm q} \propto R^{7/2}$. The fit is excellent, with
the exception of the point corresponding to the longest event. The
qualitative agreement with the theoretical expectation is striking, although
by substituting the appropriate values for the mass and accretion rate
we find that our best fit predicts rather small values of $\alpha_2$
(0.004 and 0.05 for the Schwarzschild and extreme Kerr cases
respectively).
This indicates a small viscosity in the disk, although we stress that 
$t_{visc}$ is only a time scale, so that additional corrections might
be necessary in order to allow a precise quantitative comparison.

\begin{figure}[htb]
\epsfysize=9cm
\vspace{11pt}
\hspace{0.0cm}\epsfbox{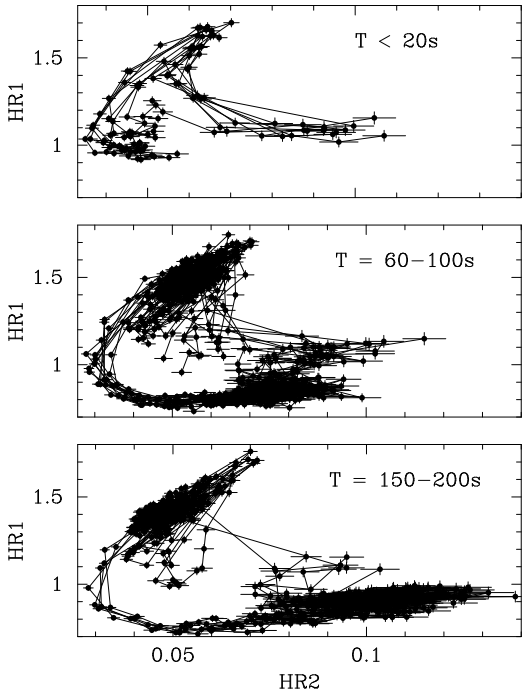}%
\caption{Color-color diagrams for events of different duration. The `low'
	intervals are correspond to the populated clouds of points on 
	the low part of the diagrams. The sequence `low'-`high'-'low'
	moves clockwise}
\label{fig:3}
\end{figure}

Let us follow an event and describe it in terms of the model.
The accretion rate $\dot M_0$ provided by the secondary is constant
(within one observation).
At the start of a quiescent period, the disk has a central hole, whose
radius is R$_{max}$. 
Outside the hole, $\dot M_0$ lies on a stable branch in the $\dot M-\Sigma$ 
curve (see \cite{tmb97a}). At all radii inside the hole, $\dot M_0$ lies on the
unstable branch: the disk lies on the lower (stable) branch. This means that
the hole does not have to be empty, but rather filled with
gas whose radiation is too soft to be detected. 
Slowly the hole in the disk is re-filled
by a steady accretion rate $\dot M_0$ from outside. Each annulus of the disk
will move along the lower branch of its S-curve in the $\dot M-\Sigma$ plane 
trying to reach $\dot M_0$.
The local density at each radius increases as the annulus moves towards the 
unstable point at a speed determined by the local viscous time scale. 
During this period, no changes are observed in the radius of the hole,
since all the matter inside does not radiate in the PCA band. In the XTE
data the disk appears to have a roughly constant radius.
The observed accretion rate during this phase is $\dot M_0$ since it is
determined from the spectrum of the radiation coming from stable regions.
At the end of this phase, 
one of the annuli will reach the unstable point and switch to the high-$\dot M$
state, where the accretion rate is larger than $\dot M_0$, causing a 
chain-reaction that will ``switch on'' the inner disk. The
observed accretion rate is now higher than the external value $\dot M_0$ since
it is determined from the spectrum of radiation originating from the unstable 
part of the disk, through which a higher $\dot M$ is flowing.
A smaller, hot radius is now observed. At the end of the outburst, the inner
disk runs out of fuel and switches off, either jumping back to the 
$\dot M < \dot M_0$ state or possibly emptying completely. 
A new hole in the disk is formed and a new cycle starts.
Notice that in this scenario the more ``normal'' state for the source
is the one at high count rates, where the disk extends all the way to the
innermost stable orbit:
in this state the energy spectrum 
is similar to that of conventional black-hole candidates
(see \cite{tl95}). 

\begin{figure*}[htb]
\epsfysize=8cm
\vspace{12pt}
\hspace{2.0cm}\epsfbox{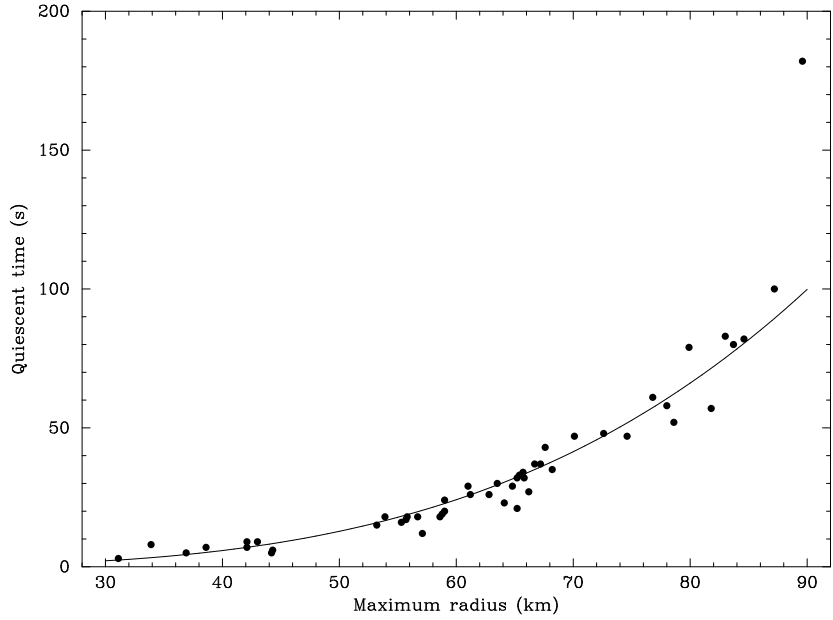}%
\caption{Correlation between the total length of an event and maximum inner
radius of the disk. The line is best fit with a power law with fixed index
$\Gamma$=3.5.}

\label{fig:2}
\end{figure*}

Not only the start and end of a major burst, but also all the small
amplitude oscillations within a burst show the same timing signature
of decaying faster than they rise. This is in agreement with what was already
noticed in \cite{tmb97a}: the rise time is determined by the speed at which
a heating wave moves through the central disk, while the faster decay time
is due to the rapid fall of matter into the black hole (or ejection into a
relativistic jet).

It has been found that when the hardness ratio HR2 exceeds 0.1, 
the power density spectrum is similar to the one observed
in black-hole transient systems during the Very High State (see \cite{vdk95}).
In these occasions, a strong 1-6 Hz QPO peak was found, positively 
correlated with the count rate \cite{cst97}.
The limit HR2$>$0.1 is an indication that the source was
in a quiescent state. Our spectral results show that 
the fast timing features (both QPO and band-limited noise, see \cite{mgr97}) 
cannot originate from the innermost
regions of the optically thick accretion disk, 
since those are missing during the quiescent phases.
The fact that the QPO frequency increases with count rate is in qualitative
agreement with the model, since a higher count rate indicates a smaller
inner disk radius, and therefore shorter time scales. 
However, the QPO seems to be associated to the power law spectral component
and not from the disk component, making this phenomenology very difficult
to understand.

The radii for the disappearing region of the disk found here are
considerably smaller than that reported in \cite{tmb97a} from an
observation where the length of the events was substantially longer. This is
entirely due to the improved knowledge of the spectral response of the
PCA. Interestingly,the length of the quiescent period and the maximum 
inner radius observed in \cite{tmb97a} are in agreement with the curve
in Figure 3 when the time is properly re-normalized to take into account
the difference in accretion rate (see Equation 1).

Our model reduces the complication of the spectral evolution of GRS~1915+105
to one parameter: the radius of the missing section of the accretion disk.
The structure of the variability is however yet to be explained.
The question to be answered is: what determines the length of the
next outburst? The model outlined above does not provide an answer, but
allows us to reformulate the question in more physical terms.
What determines how large the next missing section of the disk will be? 
In some observations the
events are very regular, in some others they seem to be random, and in
some others no events are observed at all. 
The latter might be part of extremely
long quiescent intervals. In the framework of this model, it is clear that
a regular quasi-periodic structure of events arises naturally if the
radii of the holes are all similar.
In the observation reported
here a striking one-to-one relation between quiescent and burst time is
observed \cite{tmb97b}, a relation which applies to the burst and i
quiescent states of the observation presented in \cite{tmb97a} but
is obviously not satisfied in other observations (see e.g. \cite{tcs97}) 
nor during the ``flare'' state of \cite{tmb97a}.
Moreover, as already mentioned, a few observations among the ones we analyzed
does not fit this pattern and requires a different interpretation 
(1996 June 16th is a template). Both the timing structure and the spectral
evolution in these observation is radically different from what presented
here. Nevertheless, our model provides a satisfactory interpretation of the 
cause of the changes in the X-ray emission. 

GRS~1915+105 is a remarkable 
X-ray source. Despite it 
uniqueness (and because of it), we have the chance to learn something
fundamental about accretion disks around black holes. Some of its
characteristics (the X-ray variability being the major one) are indeed
peculiar, but others (the band-limited noise and QPO in power spectra,
the thermal disk and the hard power law components in the energy spectra)
are remarkably similar to many `normal' black hole candidates. Whatever
the different states in these sources are, so far we have been observing
them on long time scales (years for persistent sources like Cyg X--1 and
GX 339--4, months for black hole transients). Here we can observe
more spectral and timing variations within one hour of observation.
In addition, coordinated observations in different bands are beginning
to provide us with the first links between what happens in the accretion
disk (observed in X-rays) and the ejection of relativistic jets (observed
in IR and radio) \cite{pfen97,eike98,mira98}.


\begin{thebibliography}{9}
\bibitem{miro94} Mirabel I.F., \& Rodr\'\i guez L.F., 1994, Nature, 371, 46.
\bibitem{bail95} Bailyn C.D., et al., 1995, Nature, 378, 157.
\bibitem{ctbl92} Castro-Tirado A.J., Brandt S., \& Lund S., 1992, IAUC 5590.
\bibitem{gmr96} Greiner J., Morgan E., \& Remillard R.A., 1996, ApJ, 473, L107.
\bibitem{mgr97} Morgan E., Remillard R.A., \& Greiner J., 1997, ApJ, 482, 993.
\bibitem{cst97} Chen X., Swank J.H., \& Taam R.E., 1997, ApJ, 477, L41.
\bibitem{tmb97a} Belloni T., et al., 1997a, ApJ, 479, L145.
\bibitem{tmb97b} Belloni T., et al., 1997b, ApJ, 488, L109.
\bibitem{tcs97} Taam R.E, Chen X., \& Swank J.H., 1997, ApJ, 485, L83.
\bibitem{fkr92} Frank J., King A., \& Raine D., 1992, ``Accretion power
	in Astrophysics'', Cambridge Univ. Press, Cambridge.
\bibitem{tl95} Tanaka Y., \& Lewin W.H.G., 1995, in ``X-ray binaries'',
	eds. Lewin W.H.G., Van Paradijs J., \& Van den Heuvel E.P.J., 
        Cambridge Univ. Press, Cambridge.
\bibitem{vdk95} Van der Klis M., 1995, in ``X-ray binaries'',
	eds. Lewin W.H.G., Van Paradijs J., \& Van den Heuvel E.P.J., 
        Cambridge Univ. Press, Cambridge.
\bibitem{pfen97} Pooley G.G., \& Fender R.P., 1997, MNRAS, 292, 925.
\bibitem{eike98} Eikenberry S.S., et al., 1998, ApJ, in press.
\bibitem{mira98} Mirabel I.F., et al., 1998, A\&A, 330, L9.

\end{thebibliography}
\end{document}